\documentclass{PoS}
\usepackage{subfig}

\title{Using INTEGRAL/SPI to study the Sun}

\ShortTitle{Using INTEGRAL/SPI to study the Sun}

\author{\speaker{R.~Rodr\'{\i}guez-Gas\'{e}n}\\
       CSNSM, IN2P3-CNRS, Univ. Paris-Sud,  \& LESIA, Observatoire de Paris, CNRS, UPMC, Univ. Paris-Diderot, France\\
        E-mail: \email{rosa.rodriguez@obspm.fr}}

\author{J.~Kiener\\
        CSNSM, IN2P3-CNRS, Univ. Paris-Sud, France\\
        E-mail: \email{jurgen.kiener@csnsm.in2p3.fr}}
       
\author{V.~Tatischeff\\
        CSNSM, IN2P3-CNRS, Univ. Paris-Sud, France\\
        E-mail: \email{vincent.tatischeff@csnsm.in2p3.fr}}

\author{C.~Hamadache\\
        CSNSM, IN2P3-CNRS, Univ. Paris-Sud, France\\
        E-mail: \email{clarisse.hamadache@csnsm.in2p3.fr}}
        
\author{K.-L.~Klein\\
	LESIA, Observatoire de Paris, CNRS, UPMC, Univ. Paris-Diderot, France\\
	E-mail:\email{ludwig.klein@obspm.fr}}
	
\author{N.~Vilmer\\
	LESIA, Observatoire de Paris, CNRS, UPMC, Univ. Paris-Diderot, France\\
	E-mail:\email{nicole.vilmer@obspm.fr}}
	
\author{and the SEPServer Consortium}

\abstract{Solar energetic particle (SEP) events are a prime opportunity to probe astrophysical particle acceleration by combining in-situ measurements of particles in space with remote sensing of the hard X-ray (HXR) and $\gamma$-ray emitting particles in the solar atmosphere. For the space environment SEPs are among the biggest hazards, particularly for spacecraft and human exploration of the solar system. Unfortunately our current understanding of the acceleration and transport of SEPs is weak. With the increasing onset of the space age it is vitally important that we increase our knowledge of the physical mechanisms related to SEPs. In this direction, the SEPServer project aims at building an on-line server that will provide the space research community with SEP data and related observations of solar electromagnetic emission. INTEGRAL/SPI acquires observations from many X-ray/$\gamma$-ray flares associated with SEP events. To make this data accessible for SEP studies we have undertaken a major effort to specify the solar observing conditions of INTEGRAL through Monte-Carlo simulations of its ACS/BGO-detector response for several flares associated with selected SEP events.}

\FullConference{An INTEGRAL view of the high-energy sky (the first 10 years)"
9th INTEGRAL Workshop and celebration of the 10th anniversary of the launch,\\
		October 15-19, 2012\\
		Bibliotheque Nationale de France, Paris, France}

\begin{document}

\section{Introduction}\label{s1}
The Sun is one of the nearest examples of an astrophysical particle accelerator. Flare-accelerated electrons and ions interacting with the ambient solar plasma produce bremsstrahlung HXR/$\gamma$-ray continuum \cite{Holman11} and $\gamma$-ray line emission \cite{Ramaty86,Vilmer03}, respectively. Quantitative information about the parent electrons and ions distributions can be inferred from HXR \cite{Kontar11} and $\gamma$-ray line measurements \cite{Vilmer11}, which provide strong constraints on the acceleration mechanisms and interaction sites.

Transient enhancements of energetic particle intensities of solar origin in space are called SEP events. The analysis of SEP events requires a comprehensive set of EM data to infer information about the origin, the acceleration processes and the transport of these particles. Diagnostics to analyse the source of energetic particles include remote sensing of their EM emission in the solar atmosphere (mainly $\gamma$-rays, HXR and radio waves), and direct measurements of particles escaping to space. In the solar context, the term energetic particle refers to electrons of energies up to 100~MeV, and protons and ions up to some GeV, which should be compared to typical thermal energies of the seed population of the order of 100~eV in the corona. The possibility of many complementary tools, including also imaging of the plasma around the presumable acceleration regions, lends a particular interest to SEPs for the investigation of particle acceleration and transport in astrophysical environments.

The work presented here is carried out within the European FP7 project SEPServer, and it aims at the study of a set of SEP events which occurred during the 23rd solar cycle (1996-2006). We are recollecting HXR and $\gamma$-ray data recorded by INTEGRAL \cite{Winkler03} spacecraft, and we are providing information about the photon energy to be detected for each solar flare. 

\section{Observing solar flares with INTEGRAL}\label{s2}

Although INTEGRAL was not designed for solar flare studies, it also observes photons of solar origin. It was designed so that the Sun is typically to the rear of the instrument, and solar photons enter the instrument through the backside. Then, the BGO-detector of the Anti-Coincidence shield (ACS) of SPI can be reached by solar high energy photons. In fact, the BGO detector is one of the largest detectors of astrophysical high energy photons presently in orbit and can be used to derive light curves without spectral information. ACS provides count rates with 50 ms time resolution above some photon energy threshold in the range of several tens of keV. The detailed value of this threshold depends on the viewing angle between the spacecraft axis and the Sun, as well as on the incident photon spectrum. This information must be derived from the modelling of the instrument. 

Such modelling is done here by means of Monte-Carlo simulations, where an evaluation of the observation conditions and the detector response of ACS/BGO-detector for each individual event has been conducted. From the Monte-Carlo simulations, with an initial photon injection which follows a power law, we have computed the effective and bulk energy, and the effective area (see below) that would characterize every event.

\subsection{Data selection}\label{s2.1}

SEPServer relies on different lists of SEP events published in the solar literature or established by the consortium. Based on the list published by \cite{Laurenza09}, Table~\ref{table1} gives the characteristics of those events where INTEGRAL/ACS detected a solar burst, as well as the results obtained from the Monte-Carlo simulations. From left to right, columns display (1) the date of the flare occurrence, (2) its peak time, (3) the flare class, all derived from GOES soft X-ray (SXR) measurements, and (4) its location. Columns (5)-(9) list parameters derived from the ACS detector simulations that will be presented in the following subsection. 

\vspace{-0.2cm}
\begin{center}
\begin{table}[h!]
\begin{scriptsize}
\caption{List of solar flares. (1) date; (2) peak time; (3) SXR class; (4)  H$_{\alpha}$ Location; (5) azimuthal angle; (6) meridional angle; (7) effective threshold energy; (8) bulk energy; (9) effective area.}
\label{table1}
\begin{tabular}{ccccccccc}
\hline 
Date & Time [UT] & Class & H$_{\alpha}$& $\theta_{I}$ [$^{\circ}$] & $\varphi_{I}$ [$^{\circ}$] & ${E}_{\rm eff}$ [keV] & ${E}_{b}$ [keV] & ${A}_{\rm eff}$ ($>$~50~keV) [cm$^{2}$]\\
\hline
\textbf{2002} & & & & & & & & \\
09-11-2002 & 13:23 & M4.9 & S12W29 & 75.4 & 179.5 & 118.5 & 191.0  & 424\\
\textbf{2003} & & & & & & & & \\
28-05-2003 & 00:27 & X3.9 & S07W21 & 96.3 & 180.5 & 114.4 & 186.8 &  216\\
31-05-2003 & 02:24 & X1.0 & S07W65 & 121.2 & 178.4 & 77.0 & 113.1 &  79\\
17-06-2003 & 22:45 & M6.8 & S08E61 & 75.8 & 179.7 & 111.8 & 182.3  &  418\\
26-10-2003 & 18:11 & X1.4 & N02W38 & 118.9 & 181.2 & 72.7 & 100.7 & 96\\
28-10-2003 & 11:10 & X18.4S & S16E07 & 122.0 & 180.7 & 78.8 & 124.1 & 73\\
02-11-2003 & 17:25 & X9.3 & S14W56 & 75.4 & 180.6 & 118.5 & 191.0 & 423\\
04-11-2003 & 19:44 & X18.4S & S19W83 & 126.3 & 180.1 & 87.5 & 204.8 & 38\\
\textbf{2004} & & & & & & & & \\
11-04-2004 & 04:19 & M1.0 & S14W47 & 72.8 & 180.7 & 115.8 & 172.0 &  467\\
07-11-2004 & 16:06 & X2.2 & N09W17 & 100.1 & 183.1 & 113.1 & 267.7  & 163\\
10-11-2004 & 02:13 & X2.8 & N09W49 & 94.7 & 177.1 & 83.6 & 176.0 & 268\\
\textbf{2005} & & & & & & & & \\
17-01-2005 & 09:52 & X4.2 & N14W24 & 111.9 & 177.2 & 79.8 & 122.7 & 81\\
20-01-2005 & 07:00 & X7.9 & N12W58 & 118.8 & 179.2 & 76.1 & 96.1 &  90\\
13-05-2005 & 16:57 & M8.5 & N12E11 & 117.1 & 179.6 & 77.0 & 108.0 & 85\\
16-06-2005 & 20:22 & M4.3 & N09W87 & 112.9 & 179.3 & 81.6 & 118.5 & 76\\
13-07-2005 & 14:49 & M5.6 & N13W75 & 67.6 & 181.8 & 107.9 & 200.1 & 524\\
27-07-2005 & 05:01 & M3.8 & $<$~E90  & 105.9 & 180.0 & 105.5 & 241.0 & 103\\
07-09-2005 & 17:40 & X18.1 & S06E89 & 77.9 & 180.1 & 115.8 & 200.1 & 400\\
13-09-2005 & 20:04 & X1.6 & S09E05 & 115.5 & 180.8 & 80.7 & 113.1 & 86\\
\hline
\end{tabular}
\end{scriptsize}
\end{table} 
\end{center}
\vspace{-1cm}

\subsection{Observational data and Monte-Carlo simulations}\label{s2.2}

\begin{figure}[h!]
\centering
\includegraphics[scale=0.55]{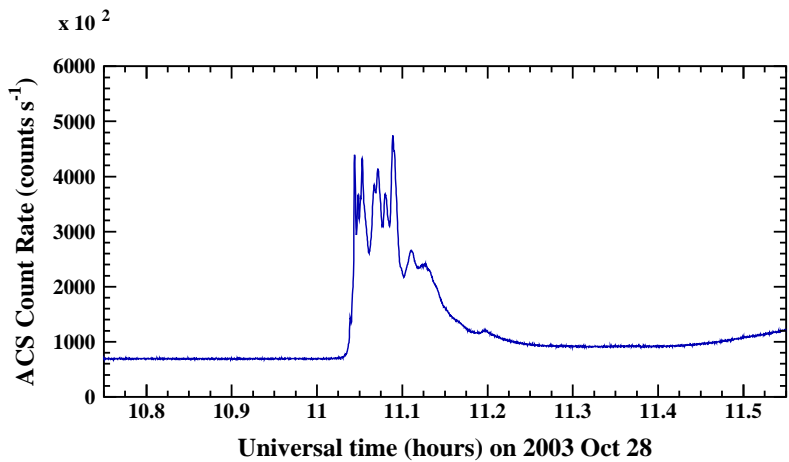}
\caption{ACS count rates data of the October 28, 2003 solar flare.}
\label{ACS}
\end{figure}

The observational data of ACS for the set of solar flares under study has been recollected. All this data will be available at \textrm{http://www.sepserver.eu}. The time resolution of the data used in the present study is 1~second and, for each solar flare, the time interval extends from 1~hour before the GOES peak time to 5~hours after. As a matter of example, Figure~\ref{ACS} shows the light curve for the October 28, 2003 solar flare. The sudden change in the registered count rates, corresponding to the high energy photons arrival coming from the solar flare is clearly seen. Such prompt increase occurs between 11.0 and 11.2~hours, in temporal correlation with the peak time in SXR that GOES recorded at 11:10~UT (11.16~hours).  

The response of ACS to incident solar high energy photons has been modelled by means of Monte-Carlo simulations, using a GEANT\footnote{~The GEANT program describes the passage of elementary particles through the matter. \textrm{http://wwwasd.web.cern.ch/wwwasd/geant/}} (CERN) based-code. To reproduce the instrument within the simulations, the detailed GSFC\footnote{~Goddard Space Flight Center, NASA: \textrm{ http://www.nasa.gov/centers/goddard/home/index.html}}  SPI computer mass model \cite{Sturner03} has been used, coupled to the INTEGRAL Mass Model (TIMM, developed at the University of Southampton, \cite{Ferguson03}) for the other parts of the spacecraft. 

\begin{figure}
\subfloat{{\includegraphics[width=0.42\textwidth]{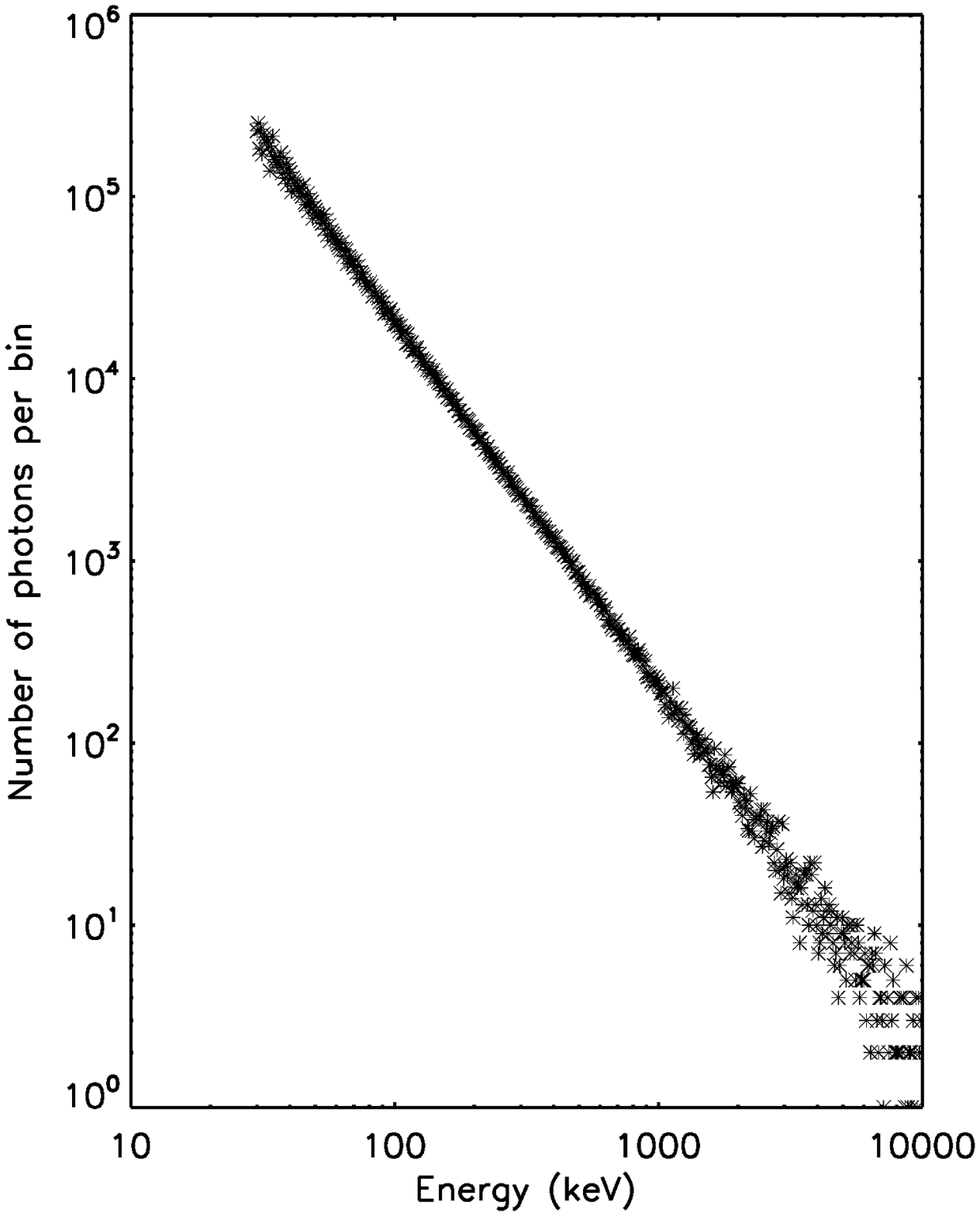}}}
\hspace{0.25cm}
\subfloat{{\includegraphics[width=0.425\textwidth]{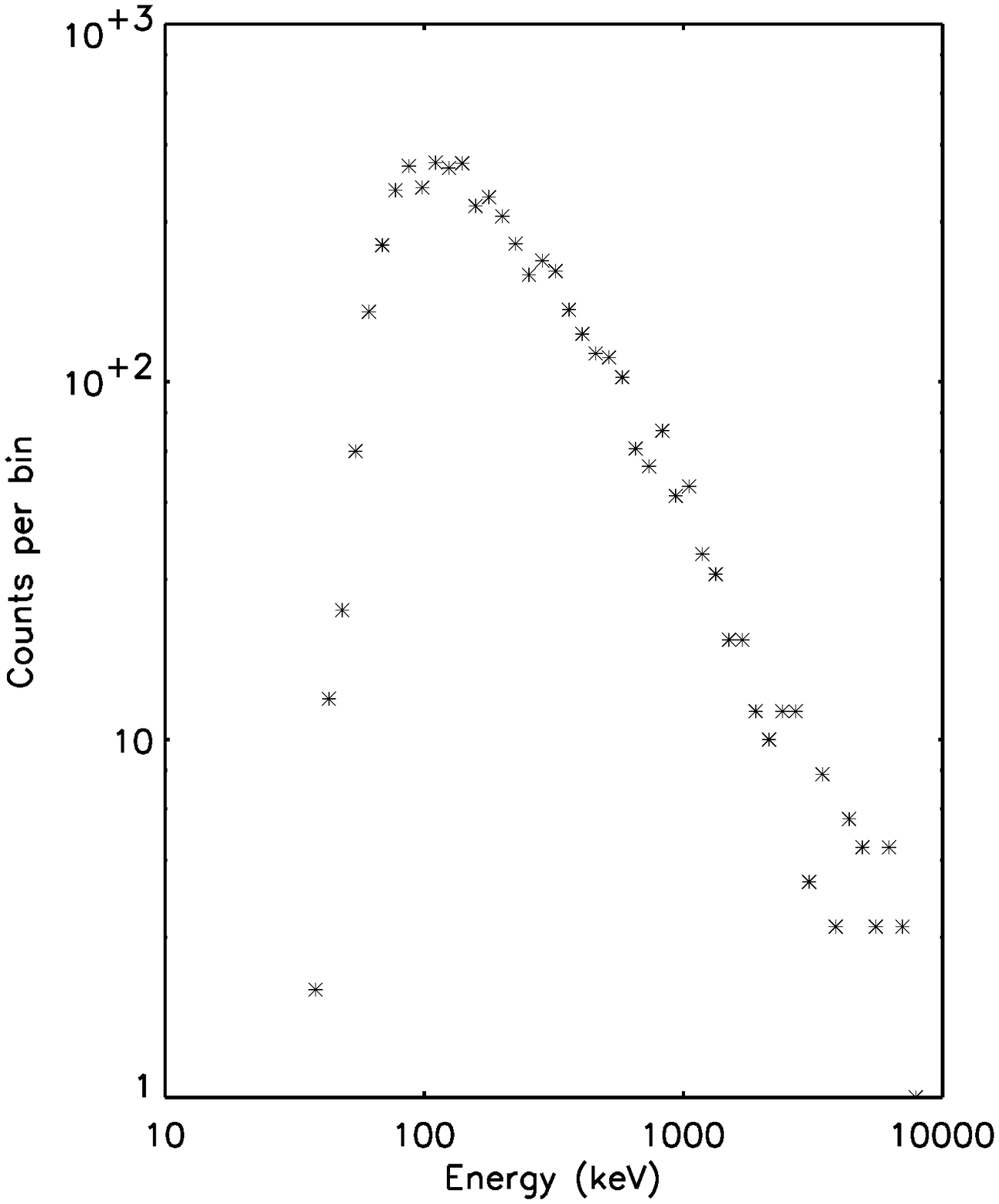}}}
\caption{Left: simulated initial power law photon injection, with $\alpha$ = 3. Right: expected response of the BGO-detector obtained from the Monte-Carlo simulations, for the scenario of the October 28, 2003 solar flare.}
\label{proves}
\end{figure}

The high energy photon injection considered follows a power law of the form: $F(E)\,\propto\,E^{-\alpha}$, setting\footnote{~The dependence of the results on the photon injection spectral index will be presented in a further paper \cite{RRG12}.} $\alpha$ = 3. The initial photon energy in the simulations is taken, randomly, within the range 30\,keV $<$ E $<$ 10~MeV. Left panel of Figure~\ref{proves} depicts the initial photon injection used for all the simulations. The photon number counting per bin has been done by means of 500 logarithmic bins within the range of the initial photon energies. The initial source is considered to be a flat disk, with a total projected area of 20106~cm$^{2}$, which injects 10$^{7}$ photons. 

For each event of the list we retrieved the orientation of INTEGRAL in space\footnote{~\textrm{http://www.isdc.unige.ch/integral/archive}}. Taking into account the spacecraft pointing direction and the exact timing of the solar flare (i.e., the position of the Sun with respect to the spacecraft at the moment of the parent activity), we computed the meridional ($\theta_{I}$) and azimuthal ($\varphi_{I}$) angles of INTEGRAL with respect to the Sun for each particular case. The computed angles\footnote{~Note that the values given for the angles are the ones used for the simulations, keeping in mind that within the Monte-Carlo code, $\varphi_{I}$ = 180$^{\circ}$ + $\varphi_{Sun}$.} are listed in columns (5) and (6) of Table~\ref{table1}. In accordance with the design specifications of INTEGRAL we found that the orientation of the spacecraft with respect to the Sun during the events covered the range 67.6$^{\circ}$ $<$ $\theta_{I}$ $<$ 126.3$^{\circ}$ in meridional angle, and 178.4$^{\circ}$ $<$ $\varphi_{I}$ $<$ 183.1$^{\circ}$ in azimuth. For the October 28, 2003 solar flare we can derive that $\theta_{I}$ = 122$^{\circ}$ and $\varphi_{I}$ = 180$^{\circ}$, in agreement with the values given by \cite{Gros04,Kiener06}.  

After reconstructing the spacecraft orientation with respect to the Sun for each solar flare, we can perform the response function simulation of the BGO-detector for each particular scenario. Then, we compute the number of hits per bin (with 50 logarithmic bins) for each case. Right panel of Figure~\ref{proves} shows, as example, the result obtained  for the October 28, 2003 solar flare scenario. The low count rates in the right part of Figure~\ref{proves} above several hundred keV are due to the binning in this particular spectrum, which was meant just as an illustration.

\subsection{Results}\label{s2.3}

Table~\ref{table1} gives the results obtained for the set of solar flares under study. Columns (5)-(9) list, respectively, the meridional and azimuthal angles of INTEGRAL with respect to the Sun used in the simulations, the computed effective threshold energy, the bulk energy, and the effective area. The bulk energy is considered as that energy for which the detector records the maximum number of hits. To determine the effective threshold energy, we have applied the criterium that the number of hits per bin will be relevant when they exceed half hits recorded at the peak. The exact value of these derived energies are dependent on the way of binning. Here we have used 500 logarithmic bins; but computations of these energies by using different kinds of binning (number of bins, constant widths ,...) have also been made. The result is a slight variation of these values, bringing us confidence on our calculations. The effective area refers to photons in the range 50\,keV\,-\,10\,MeV \cite{RRG12}.

Our calculations indicates that the effective area varies from one event to another, because the photon penetration in the detectors depends on the section of the spacecraft that they must cross: for favorable configurations (e.g., July 13, 2005 solar flare, with $\theta_{I}$ = 67.6$^{\circ}$), the effective area is bigger than for non favorable configurations (e.g., October 28, 2003 flare, with $\theta_{I}$ = 122$^{\circ}$).

Looking at the simulated response of the example shown, we deduce that ACS is sensible to photons above $\sim$\,80~keV, with the main bulk of photons coming from the energy range 80-300~keV, approximately. For incoming photons whose energy distribution follows a power-law spectrum with index $\alpha$ = 3, the maximum number of detections occurs around 125 keV. The computed total effective area is 73~cm$^{2}$. All these values account for this particular configuration, having the photons to cross the entire bottom part of the spacecraft before reaching the detector. 

\section{Summary}\label{s5}
We present part of the work that is being carried out under the FP7 SEPServer project. High energy data observed by INTEGRAL/ACS for a set of solar flares of the last solar cycle is being recollected. For all these cases, ACS recorded a substantial increase of the count rates, with a close time relation to the peak in SXR reported by GOES.

With an initial power law injection of high energy photons, and knowing the exact orientation of the spacecraft, we have performed Monte-Carlo simulations of the ACS response for each particular event. We have also derived the effective energy, the bulk energy, as well as the effective area for the energy range considered.  We find that the effective area is the quantity which varies most from event to event.

Even if INTEGRAL was not meant to perform solar observations, we sustain that its data can also be used for solar flare studies, being relevant for a better understanding of SEPs origin and of their acceleration processes. 

\acknowledgments
This research work is supported by the European Union's Seventh Framework Program (FP7/2007-2013) under grant agreement num. 262773 (SEPServer). The authors acknowledge the help provided from the INTEGRAL/SPI team and the use of the facilities of the IN2P3 computing center.

\end{document}